\documentclass[10pt]{IEEEtran}
\usepackage[a4paper, total={6in, 8in}]{geometry}
\usepackage{cite}
\usepackage{amsmath,amssymb,amsfonts}
\usepackage{graphicx}
\usepackage{textcomp}
\usepackage{xcolor}
\usepackage[inline]{enumitem}
\usepackage[normalem]{ulem} 
\usepackage{times}
\usepackage{helvet}
\usepackage{courier}
\usepackage{mathtools} 
\usepackage{booktabs}
\usepackage{makecell}
\usepackage{multirow}
\usepackage{multicol}
\usepackage{diagbox}
\usepackage{bm,tikz}
\usepackage{centernot}
\usepackage{wasysym}

\usepackage{bm}
\def\BibTeX{{\rm B\kern-.05em{\sc i\kern-.025em b}\kern-.08em
    T\kern-.1667em\lower.7ex\hbox{E}\kern-.125emX}}

\usepackage{url}
\RequirePackage{hyperref}
\PassOptionsToPackage{hyphens}{url}\usepackage{hyperref}
\hypersetup{
	colorlinks=true,
	linkcolor=blue,
	filecolor=magenta, 
	urlcolor=cyan,
	pdftitle={Overleaf Example},
	pdfpagemode=FullScreen,
}
\usepackage{xurl}
\hypersetup{pdfauthor=whatever}

\usetikzlibrary{matrix}

\usetikzlibrary{arrows,plotmarks,decorations.markings,trees,shapes}
\tikzstyle{n}=[ellipse,draw=black!100,fill=black!10,line width=.7pt,minimum width=1cm,align=center,text height=.2cm]
\tikzstyle{nodo}=[ellipse,draw=black!100,fill=black!30,line width=.7pt,minimum width=1.2cm,text width=2.2cm,align=center,minimum height=.5cm]
\tikzstyle{Qnodo}=[ellipse,draw=black!100,fill=black!10,line width=.7pt,minimum width=1.2cm,,text width=2.2cm,align=center,minimum height=.7cm]
\tikzstyle{arco}=[draw=black!80,line width=1pt, postaction={decorate}, decoration={markings,mark=at position 1.0 with {\arrow[ draw=black!80,line width=.7pt]{>}}}]
\tikzstyle{decision} = [rectangle, draw, fill=black!100,text=white, text width=4.5em, text badly centered, node distance=3cm, minimum height=3em]
\tikzstyle{block} = [rectangle, draw, fill=blue!20, text width=5em, text centered, rounded corners, minimum height=3em]
\tikzstyle{line} = [draw, -latex']
\tikzstyle{cloud} = [draw, ellipse,fill=red!20, node distance=3cm, minimum height=2em]

\def\BibTeX{{\rm B\kern-.05em{\sc i\kern-.025em b}\kern-.08em
    T\kern-.1667em\lower.7ex\hbox{E}\kern-.125emX}}
\begin{document}

\title{Modelling Assessment Rubrics through Bayesian Networks: a Pragmatic Approach\\
\thanks{This research was funded by the Swiss National Science Foundation (SNSF) under the National Research Program 77 (NRP-77) Digital Transformation (project number 407740\_187246).}
}


\author{Francesca Mangili$^1$, Giorgia Adorni$^1$, Alberto Piatti$^2$,\\  Claudio Bonesana$^1$, Alessandro Antonucci$^1$\\
$^1$Istituto Dalle Molle di Studi sull'Intelligenza Artificiale (IDSIA)\\ USI - SUPSI, Lugano, Switzerland\\
$^2$Department of Education and Learning (DFA)\\SUPSI, Lugano, Switzerland\\
{\tt \{francesca.mangili,giorgia.adorni, alberto.piatti},\\ {\tt claudio.bonesana, alessandro.antonucci\}@supsi.ch}}


\date{}

\maketitle

\begin{abstract}
Automatic assessment of learner competencies is a fundamental task in intelligent tutoring systems. An assessment rubric typically and effectively describes relevant competencies and competence levels. This paper presents an approach to deriving a learner model directly from an assessment rubric defining some (partial) ordering of competence levels. The model is based on Bayesian networks and exploits logical gates with uncertainty (often referred to as noisy gates) to reduce the number of parameters of the model, so to simplify their elicitation by experts and allow real-time inference in intelligent tutoring systems. We illustrate how the approach can be applied to automatize the human assessment of an activity developed for testing computational thinking skills. The simple elicitation of the model starting from the assessment rubric opens up the possibility of quickly automating the assessment of several tasks, making them more easily exploitable in the context of adaptive assessment tools and intelligent tutoring systems. 
\end{abstract}


\section{Introduction}

Intelligent tutoring systems (ITSs) are technological devices that support learning by interacting with the user, without the mediation of a teacher,  supplying hints and suggestions which can be effective only if calibrated to the actual user competence level. Therefore, ITSs collect data during the accomplishment of the assigned tasks, analyse the learner activities and infer its competence profile based on a predefined model of the learner skills, knowledge and behaviours, and use it to select the most appropriate intervention. The new knowledge collected along with the learning activity continuously updates the competence profile, making the interventions more focused. Therefore, a central element in developing a successful AI-based educational tool is the learner model, aiming to describe the learner by a set of hidden variables representing competencies and their relations to the observable actions performed while solving the task.  

A combination of knowledge, skills, and attitudes expressed in a context defines competencies.  
To assess them, teachers should set up realistic, authentic situations where students can apply their knowledge, skills and attitudes and compare the level of competence activated during these activities with a competence model, often specified through an assessment rubric. \cite{dawson2017assessment}. A general assessment rubric consists of a list of competence components to be assessed and a qualitative description of possible observable behaviours corresponding to different levels of such components. A rubric, therefore, describes the relationship between competencies and observable behaviours that one needs to codify in the learner model formally. 
Several sources of uncertainty and variability affect the relation between the non-observable competencies and the corresponding observable actions. A deterministic relation cannot correctly model it. Instead, probabilistic reasoning is a more appropriate approach to translating qualitative assessment rubrics into a quantitative, standardised, coherent measure of student proficiency.
In the literature, \emph{Bayesian knowledge tracing} (BKT), \emph{Item response theory} (IRT), and \emph{Bayesian networks} (BN) are all popular probabilistic approaches to model learner knowledge. 
BNs are a powerful framework for describing dependencies between skills and students' behaviours in facing complex tasks; furthermore, being graphical models highly intelligible, they are usable by experts in the elicitation of the student model. In Desmarais review of all most successful ITS experiences since Bloom seminal paper \cite{desmarais2012review}, it is acknowledged that "Probabilistic models for skill assessment are playing a key role in these advanced learning environments" and BNs are presented as the most general approach for modelling learner skills. Building on these results, we focused on BNs approaches. 

Designing a BN may require a deep understanding of BNs theory and a significant effort in eliciting the network structure and parameters or the availability of a large dataset for learning the model directly from the data. Although BN arcs can be interpreted as a causal model, their definition by experts is not always trivial due to the complexity of causal relations at play and the presence of hidden causes. On the other hand, even when the learner model structure can be accurately defined, the elicitation or learning of the BNs parameters and the computation of inferences can quickly become unmanageable. The number of parameters and the problem complexity can quickly increase with the number of arcs in the network. This issue can discourage ITS practitioners from using these tools in their applications when many skills are affecting the learner's actions. To avoid it, a solution to reduce the number of parameters in a BN-based learner model was proposed in our previous paper \cite{Anonymous2022}, which exploited the so-called \emph{noisy-OR gates} \cite{pearl1988probabilistic}. We could reduce the number of parameters to elicit from exponential to linear for the number of parents/skills. Similar advantages also concern the inference. We adopted such solution to set up a general approach for translating assessment rubrics into interpretable BN-based learner models with a complexity compatible with real-time assessment.  
To illustrate this approach, we focused on the activity proposed in \cite{piatti_2022} for the standardised assessment of algorithmic skills along the entire K-12 school path, derived two learner models with different sets of expert-elicited parameters and applied them to the dataset collected in \cite{Anonymous2022}. 
Overall, we obtain a compact and general approach to implementing a learner model given a set of competencies of interest and the corresponding assessment rubric. The resulting model has a simple structure and interpretable parameters, requiring a reasonable effort for their elicitation by experts and fast inferences allowing for real-time ITS interactions.

The paper is organised as follows: Section \ref{sec:the_model} provides some background about learner modelling based on BNs and noisy-OR gates; the approach is applied to a general assessment rubric in Section \ref{sec:ct assessment}; in Section \ref{sec:results} we illustrate the procedure on the assessment rubric developed for the CAT activity, and analyse the model inferences based on the results of the pupils observed in the CAT study \cite{piatti_2022}.    


\section{Exploiting Noisy Gates in BN-based Learner Models}\label{sec:the_model}

\subsection{BN-based learner model} \label{subsec:background}
The structure of a Bayesian network (BNs) over a set of variables is described by a directed acyclic graph \begin{math}\mathcal{G}\end{math} whose nodes are in one-to-one correspondence with the variables in the set. We call parents of a variable \begin{math}X\end{math}, according to \begin{math}\mathcal{G}\end{math}, all the variables connected directly with $X$ with an arc pointing to it. Learner models usually include a set of $n$ latent (i.e., hidden) variables \begin{math}\bm{X}:=(X_1,\ldots,X_n)\end{math} (to be called \emph{skill nodes} in the following) describing the competence profile of the learner and some $m$ manifest variables \begin{math}\bm{Y}:=(Y_1,\ldots,Y_m)\end{math} (answer nodes) describing the observable actions implemented by the learner to answer each specific task. 
Typically a bipartite structure with arcs from the skill nodes to the answer nodes, but not vice versa, is adopted.
This structure is well-suited to model assessment rubrics, resulting in a simple and interpretable set of relations modelling the fact that the presence or absence of a competence directly affects the learner's answers to questions requiring such competence. 
For this work, we focus on the case of binary skill nodes, taking value 1 if the pupil possesses the skill, and binary answers nodes, indicating whether the pupil has shown the desired behaviour or not.

\begin{figure}[htbp]
\centerline{
\begin{tikzpicture}
\node[nodo] (s1)  at (0,0) {\footnotesize (\begin{math}X_1\end{math}) Finger Mult.};
\node[nodo] (s2)  at (4,0) {\footnotesize (\begin{math}X_2\end{math}) Long Mult.};
\node[Qnodo] (q1)  at (0.,-1.) {\footnotesize (\begin{math}Y_1\end{math}) \begin{math}3 \times 4=?\end{math}};
\node[Qnodo] (q2)  at (4,-1.) {\footnotesize (\begin{math}Y_2\end{math}) \begin{math}13 \times 14=?\end{math}};
\draw[arco] (s1) -- (q1);
\draw[arco] (s2) -- (q1);
\draw[arco] (s2) -- (q2);
\end{tikzpicture}}
\caption{Example of BN-based learner model. Adapted from \cite{Anonymous2022}. }
\label{fig:bn}
\end{figure}
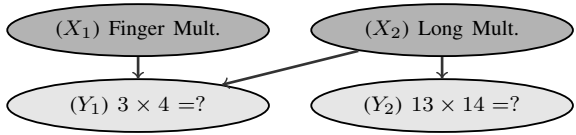

The example shown in Fig. \ref{fig:bn} graphically depicts these relations. Long multiplication skill ($X_2$) can be applied to answer both multiplications, and therefore \begin{math}X_2\end{math} is a parent node for both answer nodes \begin{math}Y_1\end{math} and \begin{math}Y_2\end{math}; instead, the multiplication in \begin{math}Y_2\end{math} cannot be computed by fingers, and thus there is no direct arc from \begin{math}X_1\end{math} to \begin{math}Y_2\end{math}. 

Once graph \begin{math}\mathcal{G}\end{math} structuring the BN is established, the definition of the BN over the $n+m$ variables \begin{math}\bm{V} = (V_1,V_2,\dots, V_{n+m})\end{math} of the network, including both  skill ($\bm{X}$) and answer ($\bm{Y}$), consists in a collection of conditional probability tables (CPTs) giving the probabilities
\begin{math}P(Y_i = 1|\mathrm{Pa}(Y_i))\end{math} that \begin{math}Y_i\end{math} takes value 1 given all possible joint states of its parent nodes \begin{math}\mathrm{Pa}(Y_i)\end{math}. 
Let $\bm{V}$ take values in $\Omega_{\bm{V}}$, the independence relations imposed from \begin{math}\mathcal{G}\end{math} by the \emph{Markov condition} 
induce a joint probability mass function over the BN variables that factorises as follows:
\begin{equation}\label{eq:joint}
P(\bm{v}=(\bm{x},\bm{y})) = \prod_{v \in \bm{v}} P(v|\mathrm{pa}(V))\,,
\end{equation}
where \begin{math}(v_1, v_2,\dots, v_{n+m})\end{math} represents a given joint state of the variable nodes in \begin{math}\bm{V}\end{math}. 
BN inference consists in the computation of queries based on Eq. \eqref{eq:joint}. In particular, we are interested in \emph{updating} tasks consisting in the computation of the (posterior) probability mass function for a single skill node \begin{math}X_q \in \bm{X}\end{math} given the observed state $y_E$ of the answer nodes $\bm{Y}_E \subseteq \bm{Y}$:
\begin{equation}\label{eq:updating}
P(x_q|\bm{y}_E) = \frac{\sum_{\bm{v}\in \Omega_{\bm{V}'}} \prod_{v\in \bm{v}} P(v|\mathrm{pa}(V))}{\sum_{\bm{v}\in \{\Omega_{\bm{V}'},X_q\}} \prod_{v\in \bm{v}} P(v|\mathrm{pa}(V))}\,,
\end{equation}
where \begin{math}\bm{V}':=\bm{V}\setminus \{\bm{Y}_E,X_q\}\end{math}. 

According to the above model, multiple parent skills may be relevant to the same answer. If the answer node \begin{math}Y_j\end{math} has \begin{math}n\end{math} parent skills, this results in \begin{math}2^n\end{math} parameters to be elicited by experts. 
Besides the elicitation effort, also the inferential complexity can become critical. \cite{Anonymous2022} discusses this and demonstrates how the use of noisy gates can avoids these issues.  
In the following section, we focus on the disjunctive noisy-OR gate, which shapes interchangeable skills and is suitable for modelling the assessment rubric of the activity from \cite{piatti_2022}. The proposed approach shares with BKT some similarities which will be emphasised later in this section, but, while the latter, in its common implementation, traces the evolution of a single skill over time, the former focuses on fine-grained skills modelling in a specific moment.

\subsection{Noisy-OR}\label{sec:noisygates}


The CPT of a \emph{noisy-OR gate} is specified as \cite{pearl1988probabilistic}:
\begin{equation}\label{eq:noisy}
P(Y_j=0|x_1,\ldots,x_n) = \prod_{i=1}^n (\mathbb{I}_{x_i=0}+\lambda_i\mathbb{I}_{x_i=1})\,,
\end{equation}
where $\lambda_{ij}>0$, $i=1,\dots,n$ are the model parameters defining the relation between $Y_j$ and its $n$ parent nodes, and  \begin{math}\mathbb{I}_A\end{math} is an indicator function returning 1 if \begin{math}A\end{math} is true and 0 otherwise.
To better understand Eq. (\ref{eq:noisy}), a typical representation of the noisy-OR networks structure, introducing \begin{math}n\end{math} auxiliary variables (also called inhibitor nodes), is shown in Fig 2. The state of \begin{math}Y_j\end{math} is deterministically imposed as the logical disjuction (OR) of the auxiliary parent nodes. This first simplification removes the need of specify the answer node CPT given the state of its parent nodes. Furthermore, we set the input variable \begin{math}X_i\end{math} as the unique parent of \begin{math}X_{i,j}'\end{math} and constrain \begin{math}X'_{i,j}\end{math} to be 0 with probability 1 when \begin{math}X_i=0\end{math}. Thus, the only parameter to be determined is \begin{math}\lambda_{i,j} = P(X_{i,j}'=0|X_i=1)\end{math}.
We can regard the auxiliary variable \begin{math}X_{i,j}'\end{math} as an \emph{inhibitor} of skill \begin{math}X_i\end{math} in performing the action described by \begin{math}Y_j\end{math}, since with probability \begin{math}\lambda_{i,j}\end{math} it makes the skill unavailable to the success of \begin{math}Y_j\end{math} even if the skill \begin{math}X_i\end{math} is indeed mastered by the learner. It can be regarded as the analogous of the slip probability in BKT models.

\begin{figure}[htp!]
\centering
\begin{tikzpicture}
\node[n] (xx1)  at (0,1) {${ X_1 }$};
\node[n] (xx2)  at (1.5,1) {${ X_2}$};
\node[] (xx3)  at (3,1) {${\ldots}$};
\node[n] (xx4)  at (4.5,1) {${ X_n}$};
\node[n] (x1)  at (0,0) {${ X_{1,j}'}$};
\node[n] (x2)  at (1.5,0) {${ X_{2,j}'}$};
\node[] (x3)  at (3,0) {${ \ldots}$};
\node[n] (x4)  at (4.5,0) {${ X_{n,j}'}$};
\node[n] (y)  at (2.25,-1) {${ Y_j}$};
\draw[arco] (xx1) -- (x1);
\draw[arco] (xx2) -- (x2);
\draw[arco] (xx4) -- (x4);
\draw[arco] (x1) -- (y);
\draw[arco] (x2) -- (y);
\draw[arco] (x3) -- (y);
\draw[arco] (x4) -- (y);
\end{tikzpicture}
\caption{A noisy gate (explicit formulation).}
\label{fig:bn2}
\end{figure}
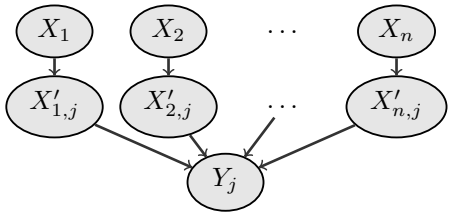

In accordance with the above description of the noisy-OR gate,  missing skill \begin{math}i\end{math} implies the inability to apply it to any question \begin{math}j\end{math}, whereas if the learner has the skill, the probability of being able to apply it depends on the specific task and is equal to \begin{math}1-\lambda_{i,j}\end{math}; the parameters of the model should therefore be related in some sense to the difficulty of the task. 
For instance, setting \begin{math}\lambda_{i,j} = 1\end{math}, implies that having skill \begin{math}X_i\end{math} has no effects on the capability of the learner to succeed in the task \begin{math}Y_j\end{math} and that the inability to answer the question \begin{math}Y_j\end{math} does not provide any information about the learner possessing skill \begin{math}X_i\end{math}. This corresponds to a missing arc in the BN graph. On the other hand, \begin{math}\lambda_{i,j} = 0\end{math} means that the presence of skill \begin{math}X_i\end{math} ensures that the learner will succeed in the task \begin{math}Y_j\end{math}. Consequently, a question of this kind would be the most informative about skill \begin{math}X_i\end{math}, especially in case of failure, since it would imply with probability 1 that the learner does not master skill \begin{math}X_i\end{math}. 

The noisy-OR can be used to describe a situation where a single skill is sufficient to answer a specific question, as in the case of Fig. \ref{fig:bn} where multiplications are solved either by fingers or by long multiplication. In that example, the relations described by that network can be modelled by a noisy-OR with four parameters, one for each skill-answer pair, with $\lambda_{1,2}=1$ to describe the fact that skill $X_1$ does not allow answering question $Y_2$.

 To apply the above model, the domain expert (e.g., the teacher) should first list the parentless skill nodes \begin{math}X_1,\dots,X_n\end{math} (with \begin{math}X_i=1\end{math} meaning that the learner possesses skill \begin{math}i\end{math}), the childless answer nodes \begin{math}Y_1,\dots,Y_m\end{math} (with \begin{math}Y_j=1\end{math} meaning a correct answer) and the parents skills relevant to each of them (setting to 1 the parameter $\lambda_{ij}$ for all non-relevant skills). Then, he should quantify for each skill relevant to \begin{math}Y_j\end{math} the value of \begin{math}\lambda_{i,j}\end{math} for a total of $n\cdot m$ parameters at most to be elicited. Finally, he should state the marginal prior probabilities $\pi_i$ of each skill, which plays the role of the initial probabilities of possessing a skill in BKT. Notice that, since the proposed approach does not model the learning process (differently from BKT which describes it through the transit probability), the concept of initial probability here represents our \emph{initial} knowledge of the learner competence profile rather than its initial level. 

\paragraph*{Leaky Models}
In a noisy-OR gate, when all skills are missing, all auxiliary variables are in the false state and, therefore, all answers must be wrong. Such model excludes the possibility of a lucky guess. To avoid this, the noisy-gates are made \emph{leaky} by adding the leak node, a Boolean variable playing the role of an auxiliary skill \begin{math}X_{j,leak}\end{math}, parent of \begin{math}Y_j\end{math} and set to \begin{math}X_{leak}=1\end{math}. Parameter $1-\lambda_{j,leak}$ describes the chances of guessing answer $Y_j$ at random, i.e., without mastering any of the relevant competences. For instance, in a multiple choice question with four options, one of which is correct, one should set \begin{math}1-\lambda_{leak}=1/4\end{math}. $1-\lambda_{leak}$ can therefore be seen as the analogous of the guess probability in BKT.

\paragraph*{Posterior Probabilities}\label{sec:posterior}
After gathering the answers from the learner, the model computes posterior inferences about the probability of the learner possing each skill. 
When the given answer is wrong, i.e., \begin{math}Y_E\end{math} is false, the noisy-OR implies that all its parent nodes \begin{math}(X_1',\ldots,X_n')\end{math} are in the false state, meaning that the learner was unable to apply any of the skill that would have led him to perform the desired action, either because the skills are indeed missing or because they were inhibited. Then, 
the posterior probability of having  \begin{math}X_q\end{math} when failing answer \begin{math}Y_E\end{math} is related to the parameter \begin{math}\lambda_{q,E}\end{math} by 
\begin{equation}\label{eq:or01}
P(X_q=1|Y_E=0) = 
\frac{\pi_q \lambda_{q,E}}{\pi_q \lambda_{q,E} + (1-\pi_q)}\,,
\end{equation}
implying a reduction in the probability that the student has the competence, the smaller the greater the inhibition probability $\lambda_{q,E}$. 
When instead the answer is correct, i.e., \begin{math}Y_{E}=1 \end{math}, it is not possible to directly propagate the evidence to the auxiliary skill nodes since it can only be stated that the learner was able to apply at least one of the competences relevant to \begin{math}Y_{E} \end{math}. In this case, $P(X_q = 1|Y_E=1)$ depends also on the parameters \begin{math}\lambda_{j,E}\end{math} assigned to the other parent skills \begin{math}X_j\end{math} as follows:  
\begin{equation}\label{eq:post2}
\begin{multlined}
P(X_q = 1|Y_E=1) = \\
=\frac{\pi_q-
\pi_q \lambda_{q,E}
\prod_{j\neq q} (1 - \pi_j (1-\lambda_{j,E}))
}{1-\prod_{j=1}^n (1 - \pi_j (1-\lambda_{j,E}))}\,.
\end{multlined}
\end{equation}
This implies an increased probability that the student has the competence, which is smaller the greater $\lambda_{q,E}$, since a skill with large inhibition probability is not likely to be the one which enabled the success in the task. 

 \section{Translating Assessment Rubric into BNs}\label{sec:ct assessment}

In this paper, we consider only assessment methods based on a task-specific assessment rubric \cite{jonsson2007use} defined for assessing a given competence through a given task or family of similar tasks. IT consists of a two-entry table with rows corresponding to a component of the given competence, described in the light of the given task, and columns describing the competence levels, from initial to complete mastery of the competence component in the given task. During the solution of the task, for each combination of component and level, a qualitative description of the behaviour expected from a person with the given level in the given component is defined. Assessing a person's skill through a task-specific assessment rubric consists of matching the behaviours of the person solving the given task (or a battery of similar tasks) with those described in the assessment rubric to identify his/her competence level in each competence component. Each cell in the rubric is called competence level. 

We define an ordering between competence levels by considering a competence level higher than another if the former is sufficient to perform all tasks requiring the latter. Each column of an assessment rubric represents the competence levels in increasing order. Therefore, all the competence levels are higher than those on their left. As shown in the CAT example below, when the competence components are conceptually overlapping, a hierarchical ordering between the rubric rows is also possible.

The competence level matching the student's behaviours may not match his/her actual level, e.g., if he is under-performing. When the task is composed of similar sub-tasks sharing the same assessment rubric, it is, therefore, possible to observe behaviours matching different competence levels in the different sub-tasks. This uncertainty is handled using the BN-based approach described in Section \ref{sec:the_model} to describe learner behaviours according to the defined assessment rubric probabilistically. Consider an assessment rubric with \begin{math}R\end{math} rows and \begin{math}C\end{math} columns. We define $R\cdot C$ hidden competence variables $X_{rc}$ taking value $1$ if the student masters the competence level corresponding to the assessment rubric's cell $(r,c)$. For each task $t$ (in a battery of $T$ similar tasks) and each competence variable, we define an observable binary variable \begin{math}Y^t_{rc}\end{math}, taking value 1 if the behaviour described in the cell $(r,c)$ is observed while solving task $t$. 

Variables $X_{rc}$ and $Y^t_{rc}$ represent, respectively, the skill and answer nodes of  the network described in Section \ref{sec:the_model}. To account for the partial ordering defined by the assessment rubric, we set as parent nodes of each answer node $Y^t_{rc}$ the skill node $X_{rc}$ corresponding to the same cell of the assessment rubric, together with all other skill nodes referring to higher competence levels. 
This structure assumes that an observed behaviour can be explained by the student mastering the corresponding competence level or a higher one. 
\section{A Case Study on K-12 Algorithmic Skills} \label{sec:results}

\begin{figure*}
	\centering
	\includegraphics[width=\linewidth]{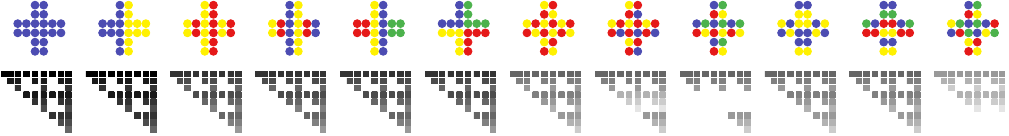}
	\caption{Cross array schemes (top) and corresponding parameters (bottom). The rows represent answers, columns the skills, and darker shades of grey lower skill-answer inhibition probabilities (white for non-relevant skills).}
	\label{fig:its}
	
\end{figure*}

To illustrate our method, we consider the Cross Array Task (CAT), an unplugged activity to assess the algorithmic skills of pupils between 3 and 16 years of age \cite{piatti_2022}.
Pupils are given a coloured cross array (Figure \ref{fig:its}, top) and asked to provide the teacher instructions to reproduce the same colouring pattern. Different levels of complexity characterise the schemes. If challenged, pupils have two aids at their disposal: an empty CAT scheme (S) at which they can point to illustrate their instruction through gestures, and feedback (F), i.e., they can see how the other person is colouring the scheme.
The instructions provided by each pupil, called an algorithm, are classified into three hierarchical categories. 
(1) \textit{0D}: 0-dimensional algorithms are based on the use of Color-One-Dot (COD) operations only.
(2) \textit{1D}, structures such as rows, diagonal, squares etc. are also used.  
(3) \textit{2D}, repetitions (loops) on dots or structures are also used.
The complexity of the produced algorithms defines the competence components of the assessment rubric. The tools used to accomplish the task determine the rubric's levels (columns). The maximum level (V) is achieved when providing instructions by voice only, without seeing the scheme being coloured; level (VS) when requesting the help of the empty scheme; level (VSF) when also asking for the feedback. We have, thus, defined a CAT assessment rubric with three rows and three columns. Furthermore, a CAT score, ranging from 0 to 4 (Table \ref{tab:catscore}), was defined to summarise the performance of a pupil in a single scheme \cite{piatti_2022}. 

\begin{table}[htb]
\caption{Definition of the CAT-score metric.\label{tab:catscore}}
\begin{center}
\begin{tabular}{l|ccc}
 & \textbf{VSF} & \textbf{VS} & \textbf{V} \\
\midrule
\textbf{0D} & 0 & 1 & 2 \\
\textbf{1D} & 1 & 2 & 3 \\
\textbf{2D} & 2 & 3 & 4 \\
\end{tabular}
\end{center}
\end{table}

Besides ordering the levels in the columns of the rubric, we define an ordering also on the rows, since mastering algorithms of higher complexity implies mastering simpler ones. Therefore, we can say that competence levels $X_{rc}$ are higher than $X_{r'c'}$ whenever $c>c'$ and $r>= r'$, or $c=c'$ and $r>r'$. When, instead, $c>c'$ but $r<r'$, neither skill can be said to dominate the other. 

As described in the previous section, all competence levels in the rubric are assigned both a hidden variable $X_{rc}$ (skill nodes) and an observable variable $Y^t_{rc}$ (answer nodes) for each task $t= 1,\dots,12$. Observing \begin{math}Y^t_{rc} = 1\end{math} means that the pupil has solved the CAT \begin{math}t\end{math} using an algorithm of complexity corresponding to the $c$-th row of the rubric, and asking the help admitted by the $r$-th column. As an example, \begin{math}Y^k_{11}=1\end{math} means that the pupil has solved the $k$-th scheme by a 0D algorithm, using all helps (VSF).
Theoretically, all answer nodes should be observed (or observable) through specific interactions with the learner. However, the dataset in \cite{piatti_2022}, here used to test the proposed approach, was collected by proposing CATs to pupils and letting them choose the algorithm and the help they wished to use. We, therefore, encoded the collected answers as follows, to make such a dataset compatible with our model. A task $t$ solved at level $c^*$ by an algorithm with complexity $r^*$ was translated into $Y^t_{rc} = 1$ for all competence levels $rc$ lower than or equal to $r^*c^*$, thus assuming that the pupil would have been able (if requested) to implement solutions requiring a lower competence level; similarly, we set $Y^t_{rc} = 0$ for all higher levels while leaving non-comparable answer nodes unobserved.

\begin{table*}[ht]
\centering
\caption{Posterior probabilities \begin{math}P(X_{rc}=1|\bm{y}^{(j)})\end{math} of model 1/model 2 for three representative pupils.\label{tab:posteriors}}
\begin{tabular}{c|ccccccccc}
\toprule
 ~ &  \multicolumn{9}{c}{\textbf{\begin{math}P(X_{rc}=1|\bm{y}^{(j)})\end{math}}}\\ 
\textbf{Student}&{\footnotesize \begin{math}X_{11} \end{math}}&{\footnotesize \begin{math}X_{12} \end{math}}&{\footnotesize \begin{math}X_{13} \end{math}}&{\footnotesize \begin{math}X_{21} \end{math}}&{\footnotesize \begin{math}X_{22}\end{math}}&{\footnotesize \begin{math}X_{23}\end{math}}&{\footnotesize \begin{math}X_{31}\end{math}}&{\footnotesize \begin{math}X_{32}\end{math}}&{\footnotesize \begin{math}X_{33}\end{math}}\\ 
$j$&{\footnotesize 0D-VSF}&{\footnotesize 0D-VS}&{\footnotesize 0D-V}&{\footnotesize 1D-VSF}&{\footnotesize 1D-VS}&{\footnotesize 1D-V}&{\footnotesize 2D-VSF}&{\footnotesize 2D-VS}&{\footnotesize 2D-V}\\ 
\midrule
1 & 0.54/0.55	&	0.70/0.76	&	0.92/0.98	&	0.64/0.66	&	0.71/0.89	&	0.17/0.99	&	0.71/0.74	&	0.21/0.79	&	0.00/0.05 \\ 
21 &  0.54/0.55	&	0.70/0.76	&	0.99/1.00	&	0.69/0.74	&	0.97/0.99	&	1.00/1.00	&	0.89/0.93	&	1.00/1.00	&	1.00/1.00	\\ 
75 &  0.54/0.55	&	0.31/0.50	&	0.00/0.01	&	0.64/0.66	&	0.00/0.04	&	0.00/0.00	&	0.07/0.19	&	0.00/0.00	&	0.00/0.00\\ 
\bottomrule
\end{tabular}
\end{table*}

Concerning parameters' elicitation, uniform prior probabilities, i.e., \begin{math} \pi_{rc} = 0.5\end{math}, are assigned to each skill, while two sets of values are considered for the inhibition probabilities \begin{math}\lambda^t_{i, j}\end{math}.
The first one, hereafter referred to as Model 1, is very basic, as it assigns the same value, namely \begin{math}\lambda^t_{i,j} = 0.2 \end{math}, to all parameters, except those corresponding to skills non-relevant to answer \begin{math}Y^t_{i,j}\end{math}, in which case \begin{math}\lambda^t_{i,j} = 1 \end{math}. The goal of this model is to analyse the effect of the constraints resulting from the ordering of the skills alone. 
The second, called Model 2, aims at describing more in detail the difficulty that the pupils can encounter in applying their skills to different schemes. Left aside non-relevant skills for which \begin{math}\lambda^t_{i,j}\end{math} remains equal to 1, the strength of the remaining skill-question relation was given ten possible levels, namely, \begin{math}\lambda^t_{i,j} \in 1- \{0.45, 0.50, 0.55, 0.60, 0.65, 0.70, 0.75, 0.80, 0.85, 0.90\}\end{math} depending on the characteristics of each CAT scheme. Our succinct elicitation setup allows summarizing graphically both the BN topology and its parameter values at the bottom of Figure \ref{fig:its}. For both models, a leak node with guess probability \begin{math}1 - \lambda_{\text{leak}} = 0.1\end{math} has been associated to all answer nodes. We implemented the underlying BN within the CREMA Java library \cite{huber2020a}.

\section{Results} \label{sec:results2}

We considered the responses provided by all 109 pupils included in the study by Piatti et al. \cite{piatti_2022}, calculated the student's average CAT scores over the twelve administered CAT schemes and compared them to probabilistic CAT scores computed for the BN-based models and defined as the expected number of competence levels mastered by the student, which corresponds to the sum of the marginal posterior probabilities of all skill nodes. The correlation between the original CAT score and the probabilistic scores is high (Pearson correlation coefficient of 0.94 for both BN-models) showing the consistency between the probabilistic assessment and the one by experts. The BN-based models, however, provide more detailed information about student competence profiles in the form of posterior probabilities for each competence level. 

The probabilistic scores of the two BN-based models defined are similar. However, it is possible to discern relevant differences in the posterior probabilities of the individual skill nodes. To show this and to demonstrate the interpretability of the model results, we show in Table \ref{tab:posteriors} hereafter the answers provided by three representative pupils and the corresponding posterior probabilities inferred by the models. 
Less straightforward is the situation of pupil 1 which was unsuccessful in using 1D algorithms with voice (1D-V) in schemes 3, 8, 10, 11, and 12 but was successful in all other schemes. Then, according to Model 1, which weights all schemes equally, he is not proficient in the 1D-V skill but only in the 1D-VS one (i.e., e must be supported by the empty scheme to produce a successfull 1D algorithm). Model 2 grants a larger probability to the 1D-V skill since it assigns larger inhibition probability to the 1D-V skill nodes when the task is more difficult. Therefore, failing to apply the 1D-V strategy to more difficult scheme does not reflect a lack of this skill if it is, instead, correctly implemented in easier tasks. 
Pupil 75 usually achieves medium to low-level solutions using 1D or 0D VS skills. He never approaches the problem using higher-level VS skills. Since 1D-VS fails on easy schemes, both models assign non-negligible probabilities to 0D-VS and 1D-VSF. 
Notice that, since low competence levels are associated with fewer answer nodes, their presence is less accurately assessed  by the administered tasks than that of higher levels.

\section{Conclusions}
In this work, we have proposed a procedure for deriving a learner model for automatic skill assessment directly from the competence rubric of any set of tasks. 
The approach has been illustrated and its feasibility demonstrated
by implementing the automatic assessment of the
cross-array task \cite{piatti_2022}. Results show that the derived model and its inferences can be easily interpreted. The model will
be used in an application for the adaptive assessment of
pupil computational skills.

 The merits of the approach are the simple elicitation, and the fast inferences resulting from the use of nosy-OR BNs \cite{Anonymous2022}. Its limitations are mainly two: either disjunctive or conjunctive gates must be used, while sometimes it would be useful to combine both or implement more general relations between skills; although partially implied by the structure of the model, the ordering between competence levels is not strictly enforced. Therefore, the current work extends the model to overcome these limitations while avoiding increasing the computational burden significantly.

\bibliographystyle{IEEEtran}
{\footnotesize
\bibliography{biblio}}

\end{document}